\documentstyle[aps,epsfig,psfig]{revtex}
\advance \topmargin by -\headheight
\advance \topmargin by -\headsep
\DeclareFixedFont{\xsf}{OT1}{cmss}{m}{n}{10}

\begin{document}
 
\def \lesssim{\leavevmode{\raisebox{-.5ex}{ $\stackrel{<}{\sim}$ } } }
\def \gtrsim{ \leavevmode{\raisebox{-.5ex}{ $\stackrel{>}{\sim}$ } } }
\def \subarr{  \begin{array}{c} \mbox{\scriptsize{$ \{ n_{\ell}\}$ }} \\ 
        \mbox{\scriptsize{ $\{m_{\ell}\}$}}             \\ 
        \mbox{\scriptsize{ $\{p_{\ell}\}$ }}  \end{array} }
\def \subneq{\begin{array}{c} \mbox{\scriptsize{$j=1$}} \\ 
	\mbox{\scriptsize{$j\neq i$}}  \end{array} }
\def \subun{ \leavevmode{\raisebox{-1.6ex}{$\stackrel{
	\textstyle{\ell_i,\ell_j}}
	{\mbox{\scriptsize{unconstrained}}} $}} }
\def \maxj{\begin{array}{c} \mbox{max} \\ 
        \mbox{\scriptsize{ $j$}}             \\ 
        \mbox{\scriptsize{ $ 0 \leq j \leq \jmax $ }}  \end{array} }
\def \maxQz{\leavevmode{\raisebox{-1.0ex}{$\stackrel{\mbox{\scriptsize{max}}}
	{\mbox{\scriptsize{Q, $\zbar$}}} $}} }
\def \npfactori{\leavevmode{\raisebox{-.5ex}{ $\stackrel{N}{p}$ } } }
\def \Nchoosen{\begin{array}{c} N \\ n \end{array} }
\def \be{\begin{equation}}
\def \ee{\end{equation}}
\def \bea{\begin{eqnarray}}
\def \eea{\end{eqnarray}}
\def \zn{z_{\mbox{\tiny N}}}
\def \zbar{\overline{z}}
\def \Zbar{\overline{Z}}
\def \znq{z_{\mbox{\tiny NQ}}}
\def \en{\epsilon_{\mbox{\tiny N}}}
\def \dele{\delta \epsilon}
\def \delen{\delta \epsilon_{\mbox{\tiny N}}}
\def \kboltz{k_{\mbox{\tiny B}}}
\def \kf{k_{\mbox{\tiny F}}}
\def \kN{k_{\mbox{\tiny N}}}
\def \ko{k_{\mbox{\tiny 0}}}
\def \etaH{\eta_{\mbox{\tiny H}}}
\def \etaC{\eta_{\mbox{\tiny C}}}
\def \ebar{\overline{\epsilon}}
\def \Ebar{\overline{E}}
\def \qmg{Q_{\mbox{\tiny MG}}}
\def \zmg{\zbar_{\mbox{\tiny MG}}}
\def \qf{Q_{\mbox{\tiny F}}}
\def \qu{Q_{\mbox{\tiny U}}}
\def \zf{\zbar_{\mbox{\tiny F}}}
\def \qst{Q^{\star}}
\def \zst{\zbar^{\star}}
\def \delF{\Delta F^{\star}}
\def \Del{\Delta}
\def \gtrsim{ \leavevmode{\raisebox{-.5ex}{ $\stackrel{>}{\sim}$ } } }
\def \lessim{ \leavevmode{\raisebox{-.5ex}{ $\stackrel{<}{\sim}$ } } }
\def \slev{s_{\mbox{\tiny LEV}}}
\def \le{\ell_{\mbox{\tiny E}}}
\def \lc{\ell_{\mbox{\tiny C}}}
\def \lec{\ell_{\mbox{\tiny EC}}}
\def \delsq{\Delta^2}
\def \tbar{\overline{\tau}}
\def \tauo{\tau_{\mbox{\scriptsize o}}}
\def \tauL{\tau_{\mbox{\tiny L}}}
\def \tauMG{\tau_{\mbox{\tiny mg}}}
\def \tauRC{\tau_{\mbox{\tiny rc}}}
\def \SMG{S_{\mbox{\tiny MG}}}
\def \Smix{S_{\mbox{\tiny MIX}}}
\def \SP{S_{\mbox{\tiny P}}}
\def \tFbar{\overline{\tau_{\mbox{\tiny F}}}}
\def \Sqz{S\left(Q,\zbar\right)}
\def \wo{\omega_o}
\def \w{\omega}
\def \wkr{\omega_{\mbox{\tiny KR}}}
\def \qdag{q^{\neq}}
\def \qddag{q^\ddag}
\def \Qddag{Q^\ddag}
\def \Fdag{F^{\neq}}
\def \Qdag{Q^\dag}
\def \e{\mbox{e}}
\def \En{E_{\mbox{\tiny N}}}
\def \Enn{E_{\mbox{\tiny NN}}}
\def \bn{b_{\mbox{\tiny N}}}
\def \Ed{E_{\mbox{\tiny D}}}
\def \Uijn{U_{ij}^{\mbox{\tiny N}}}
\def \eij{\epsilon_{ij}}
\def \eik{\epsilon_{ik}}
\def \eijn{\epsilon_{ij}^{\mbox{\tiny N}}}
\def \Dijn{\Delta_{ij}^{\mbox{\tiny N}}}
\def \Dij{\Delta_{ij}}
\def \Eg{E_{\mbox{\tiny G}}}
\def \Ef{E_{\mbox{\tiny F}}}
\def \sQ{\sigma_{\mbox{\tiny Q}}}
\def \FQ{F_{\mbox{\tiny Q}}}
\def \SQ{S_{\mbox{\tiny Q}}}
\def \PQ{P_{\mbox{\tiny Q}}}
\def \AQ{A_{\mbox{\tiny Q}}}
\def \tQ{\tau_{\mbox{\tiny Q}}}
\def \pq{p_{\mbox{\tiny Q}}}
\def \pro{p_{\rho}}
\def \psig{p_{\sigma}}
\def \pxj{p_{x_{j}}}
\def \pyj{p_{y_{j}}}
\def \lambQ{\lambda_{\mbox{\tiny Q}}}
\def \gammaQ{\gamma_{\mbox{\tiny Q}}}
\def \Vb{V_{\beta}}
\def \Qrel{Q'}
\def \ms{m^{\star}}
\def \si{\sigma_{i}}
\def \sj{\sigma_{j}}
\def \ri{r_{i}}
\def \rj{r_{j}}
\def \Fext{F_o \mbox{e}^{i\omega t}}
\def \FextkT{\frac{F_o}{\kboltz T} \mbox{e}^{i\omega t}}
\def \peq{\rho_{\mbox{\tiny EQ}}\left(Q\right)}
\def \drho{\delta \rho \left(Q,t\right)}
\def \drhop{\delta \rho \left(Q,t'\right)}
\def \DA{\Delta_a}
\def \DB{\Delta_b}
\def \DN{\Delta_{\mbox{\tiny N}}}
\def \Fbardag{\overline{F}^{\neq}}
\def \Fbar{\overline{F}}
\def \fbarm{\overline{f}_m}
\def \Fcal{{\cal F}(m)}
\def \fbar{\overline{f}}
\def \FN{F_{\mbox{\tiny N}}}
\def \mdag{m^\dag}
\def \ei{\epsilon_i}
\def \ej{\epsilon_j}
\def \Qu{Q_{\mbox{\tiny U}}}
\def \Qsurf{Q_{\mbox{\tiny SURF}}}
\def \D{\partial}
\def \DF{\Delta F}
\def \DFdag{\Delta F^{\neq}}
\def \dij{\delta_{ij}}
\def \Qdag{Q^{\neq}}
\def \Qddag{Q^{\neq}}
\def \Qo{Q^{o}}
\def \En{E_{\mbox{\tiny N}}}
\def \Qistar{Q_i^{\star}}
\def \Ebar{\overline{E}}
\def \ebar{\overline{\epsilon}}
\def \ebarN{\overline{\epsilon}_{\mbox{\tiny N}}}
\def \en{\epsilon_{\mbox{\tiny N}}}
\def \tauF{\tau_{\mbox{\tiny F}}}
\def \lbar{\overline{\ell}}
\def \Tg{T_{\mbox{\tiny G}}}
\def \Tf{T_{\mbox{\tiny F}}}
\def \Ta{T_{\mbox{\tiny A}}}
\def \Tc{T_{\mbox{\tiny C}}}
\def \Tco{T_{\mbox{\tiny C}}^o}
\def \zc{z_c}
\def \sc{s_{\mbox{\tiny C}}}
\def \kB{k_{\mbox{\tiny B}}}
\def \wq{\omega_{\mbox{\tiny Q}}}
\def \a{\alpha}
\def \b{\beta}
\def \s{\sigma}
\def \so{\sigma_o}
\def \eps{\epsilon}
\def \fH{f_{\mbox{\tiny H}}}
\def \Ff{F_{\mbox{\tiny F}}}
\def \Fu{F_{\mbox{\tiny U}}}
\def \Thc{T_{\mbox{\tiny H-C}}}
\def \d{\delta}
\def \Fmf{F_{\mbox{\tiny MF}}}
\def \kF{k_{\mbox{\tiny F}}}
\def \Ei{E_i}
\def \Ea{E_{\mbox{\tiny A}}}
\def \Fstar{F^{\star}}
\def \dij{\delta_{ij}}
\def \di{\delta_{i}}
\def \dj{\delta_{j}}
\def \Qf{Q_{\mbox{\tiny F}}}
\def \ejstar{\ej^{\star}}
\def \iset{i_1 i_2 \cdots i_p}
\def \iless{1\leq i_1<i_2\cdots <i_p\leq M}
\def \z{\zeta}
\def \zhat{\widehat{\zeta}}
\def \chat{\widehat{c}}
\def \Dhat{\widehat{D}}
\def \sdag{\tilde{s}^{\neq}}
\def \Ne{N_e}
\def \pa{p_{\alpha}}
\def \pao{p_{\alpha}^o}
\def \pio{p_{i}^o}
\def \pb{p_{\beta}}
\def \na{n_{\alpha}}
\def \nb{n_{\beta}}
\def \setna{\{ \na \}}
\def \eab{\epsilon_{\alpha\beta}}
\def \deab{\delta\epsilon_{\alpha\beta}}
\def \Om{\Omega}
\def \bg{b_{\mbox{\tiny G}}}
\def \Estar{E^{\ast}}
\def \bstar{b^{\ast}}
\def \xg{x_{\mbox{\tiny G}}}
\def \Nchoosen{\begin{array}{c} N_{\mbox{\tiny{R}}} \\ n \end{array} }
\def \dE{\delta E}
\def \dq{\delta q}
\def \Omt{\Omega_{\mbox{\tiny T}}}
\def \PB{P_{\mbox{\tiny B}}}
\def \Fdotbar{\dot{\Fbar}}
\def \wn{w_{\mbox{\tiny N}}}
\def \PR{P_{\mbox{\tiny R}}}
\def \sb{\sigma_b}
\def \Pbuff{P_{\mbox{\tiny BUFF }}}

\renewcommand{\theequation}{\arabic{equation}}

\title{Buffed energy landscapes: Another solution to the
kinetic paradoxes of protein folding}
\author{{Steven S. Plotkin} \\
{\it Department of Physics and Astronomy, University of British
Columbia, Vancouver, Canada V6T1Z1} \\
Peter G. Wolynes \\
{\it Department of Chemistry and Biochemistry and Center for
Theoretical Biological Physics, University of California at
San Diego, La Jolla 92093}}

\maketitle

\renewcommand{\thesection}{\arabic{section}}
\setcounter{section}{0}
\renewcommand{\theequation}{\arabic{equation}}

\newcounter{saveeqn}
\newcommand{\alpheqn}{\setcounter{saveeqn}{\value{equation}}
\stepcounter{saveeqn}\setcounter{equation}{0}
\renewcommand{\theequation}
      {\mbox{\thesection.\arabic{saveeqn}\alph{equation}}}}
\newcommand{\reseteqn}{\setcounter{equation}{\value{saveeqn}}
\renewcommand{\theequation}{\arabic{equation}}}

\vspace{0.25cm}
\noindent{{\bf Abstract} }
\vspace{0.25cm}

The energy landscapes of proteins have evolved to be different from
most random heteropolymers.  Many studies have concluded that
evolutionary selection for rapid and reliable folding to a given
structure that is stable at biological temperatures leads to energy
landscapes having a single dominant basin and an overall funnel
topography. We show here that, while such a landscape topography is
indeed a sufficient condition for folding, another possibility also
exists, giving a new class of foldable sequences.  These sequences
have landscapes that are only weakly funneled in the conventional
thermodynamic sense, but have unusually low kinetic barriers for
reconfigurational motion. Traps have been specifically removed by
selection.  Here we examine the possibility of folding on these
"buffed" landscapes, by mapping the determination of statistics of
pathways for the heterogeneous nucleation processes involved in
escaping from traps to the solution of an imaginary time Schroedinger
equation. This equation is solved analytically in adiabatic and
``soft-wall'' approximations, and numerical results are shown for the general case. The
fraction of funneled vs. buffed proteins in sequence space is
estimated, suggesting the statistical dominance of the funneling
mechanism for achieving foldability.\\

\setcounter{equation}{0}

{\bf Introduction.} $\;$ The mechanisms of protein folding are manifold. This diversity can
only be captured by a global statistical survey of the protein's
energy landscape. What can be said about this landscape merely from
the knowledge that proteins fold?

A foldable protein is a heteropolymer which through Brownian motion
finds a particular 
structure within a short time, and once this structure is found,
essentially stays in that structure. This requirement is both
thermodynamic and kinetic.
Thermodynamically, the folded state of a protein
must be stable and tolerably unique. 
While conformational substates can be functionally important and are
decidedly present~\cite{FrauenfelderH91}, the specificity of proteins
required for their work in the cellular network leads to strong
constraints on the structure of the active sites and binding sites of
proteins. 

For random heteropolymers, thermodynamic foldability depends on the
relation between the physiological temperature and the glass
temperature, which depends on the amino acid composition. For some
compositions, e.g. near homopolymers, $\Tg$ is low, while for other
compositions using many chemically distinct amino acids, $\Tg$ could
be higher than physiological temperatures.
If physiological temperatures are very high, only a small fraction of
all sequences will be foldable in a thermodynamic sense, because few
sequences would have sufficient stability in the ground state
conformation against entropic costs. At sufficiently low temperatures,
all sequences would have stable ground states, providing the
solvent is such that interaction free energies do not change much as
temperature is lowered.  But in this low temperature case, the
kinetics of folding would become the controlling factor.
For a typical random sequence of amino acids, while the lowest energy
state becomes thermodynamically stable only below the glass transition
temperature $\Tg$, at these low temperatures several other {\it traps}
will also become stable. The average escape time from these traps
scales exponentially in chain length. Thus for long enough chains
folding will be slow.

One resolution of this conflict between kinetic and thermodynamic
constraints is for proteins to have evolved to have compositions with
$\Tg < T$, but to be unusual sequences that are nevertheless stable in
some configuration above $\Tg$ via the introduction of an energy gap
in the density of states, so that one energetic basin dominates. The
energy landscape is funneled, with a single dominant basin. According
to this view, proteins are atypical heteropolymers, by having their
folding transition temperature $\Tf$ above the glass temperature $\Tg$
ordinarily to be expected from the overall amino acid
composition. This is the quantitative form of the principle of minimal
frustration that was put forward by Bryngelson and
Wolynes~\cite{BryngelsonJD87}.

An extreme limit of the principle of minimal frustration would be for
there to exist perfect consistency of interactions in the native
state, an idea due to G\={o}~\cite{GoN75}. In the
space of all sequences, such perfection is less likely than
merely achieving a satisfactory level of resolved frustration. Most
microscopic potential models suggest natural proteins still exhibit
frustration in their ground state. The existence of this frustration
is buttressed by the observation that many proteins can be further
stabilized by single site mutations.  Nevertheless, the perfect funnel
model does describe the kinetics of folding of many proteins as a
first approximation, as seen by the good agreement of $\phi$ values
predicted using G\={o} models with experiments~\cite{TakadaS99:pnas}.

The iconoclasts among us, nevertheless, must ask whether the
achievement of a funneled landscape with $\Tf/\Tg >1$ is the only way
the kinetic/thermodynamic requirements for folding can be met. Rather
than finding sequences that increase the native stability under
conditions where dynamics of a random sequence would be facile, we can
imagine designing an abnormally trap-free landscape giving unusually
high mobility throughout the misfolded parts of the energy landscape.

In this case, heteropolymers having compositions giving $\Tg > T$ would
easily have a ground state that is stable, but it would be
necessary to design out the barriers for escaping the deep traps, so
that those barriers 
were atypically small. The density of low energy states would not
be atypical in this case, rather what is atypical is the size of the kinetic
barriers between the states. Such an energy landscape could be said to
be ``buffed''. Even on such a buffed energy landscape, the native
state must still have some extra stability to avoid the time-scale problems
of a random search. But the extra stability in this scenario need not
scale extensively with system size, and may scale as a non-extensive
fluctuation $\sim N^{1/2}$ or smaller.

It is intriguing that some of the successes of theories based on the
strongly funneled landscape would preserved for theories based on
buffed landscapes. For example, because the ruggedness {\it en-route}
to the native structure is reduced and the effects of transient
trapping diminished, native structure formation (as measured through
$\phi$-values) will once again be dominated by polymeric properties.
However there should be observable differences between buffed and
funneled landscapes in stability and robustness.  The native state on
a funneled landscape is marginally stable with respect to a large
ensemble of unfolded states having significant entropy. The native
state on a buffed landscape is marginally stable with respect to a few
dissimilar low energy states that are kinetically connected to
it. Thus the thermal unfolding transition on a buffed landscape should
be less cooperative than on a funneled landscape. The folded structure
and folding mechanism are also less robust on a buffed landscape,
since mutations can more easily open up or close off regions of
searchable phase space. These differences between the properties
expected for buffed sequences and natural proteins suggest that
natural protein landscapes in the main are funneled rather than
buffed.

Yet given the possibility of having a buffed energy landscape, the
explanation for the prevalence of funneled landscapes must be sought
in statistics and population biology~\cite{GovindarajanS97:prot}. The
question we address is 
whether the fraction of sequences with buffed landscapes is comparable
or larger than the fraction of sequences with minimally frustrated
landscapes in the conventional sense. This is the task of the present
paper. \\

{\bf Nucleation.} $\;$ Both folding and trap escape can be thought of as the growth of a new
phase into a pre-existing phase. This growth requires overcoming a
barrier of finite height. According to classical nucleation
theory, we can introduce a reaction coordinate $N_e$ describing the
amount of new phase, $0 \leq \Ne \leq N$, where $N$ is the 
length of the chain. The typical free energy profile as a
function of $\Ne$ is  
\be
\Fbar(\Ne) = F(0) - f \Ne + \s \Ne^z \: .
\label{fnucleus}
\ee
Here $f$ is the bulk free energy gain per residue in the new phase and
$\s$ is the 
surface free energy cost per residue. These parameters are each a function
of temperature. 

For a uniformly stable phase, the exponent $z=2/3$. However in a
sufficiently 
heterogeneous medium the interface between phases can relax to a lower
free energy configuration which is roughened. The larger the nucleus,
the more the interface can relax, and the smaller the surface tension.
For a wide class of models in $3$
dimensions in this regime~\cite{Villain85,Kirkpatrick89}, 
$
\s = \so/\Ne^{1/6}
$
and the surface free energy cost scales as $\Ne^{1/2}$ rather than
$\Ne^{2/3}$. This is also true for a random first order glass
transition. 

Setting $\D \Fbar(\Ne)/\D \Ne = 0$ gives the typical critical nucleus size,
$
\overline{\Ne}^{\neq} \equiv N \overline{n}_{\mbox{\tiny F}}^{\neq} = \left(
z\s/f\right)^{1/(1-z)} 
$
where $0 \leq n_{\mbox{\tiny F}} \leq 1$. The
typical free energy barrier height is
\be
\Fbardag = 
\s (1-z) \left( z\s/f\right)^{z/(1-z)} 
= \s
(1-z) \, n_{\mbox{\tiny F}}^{\neq \, z} N^z \: .
\label{eqfdag}
\ee
Thus the barrier arises from surface cost, and scales like $N^z$.

Generally for glassy systems at high enough temperature, trap escape
becomes a downhill process, corresponding to a vanishing of the
surface tension in the nucleation picture. \\

{\bf Atypically rugged and funneled sequences.} $\;$ Suppose each
amino acid were chosen 
from a pool with probability $\pio$ for residue type $i$, where $1 <
i <m$ with $m=20$ and $\sum_{i} \pio = 1$. 
Shorter sequences may have compositions with probabilities of occurrence $\{p_i\}$ differing from 
$\{\pio\}$, but as the chain length increases, $p_i$ approaches
$\pio$. The probability of a composition $\{n_{i}\}$, $0\leq
n_{i} \leq N$, chosen randomly from a pool with natural abundances
$\{\pio\}$ is given by a multinomial distribution. We are interested
in the likelihood of finding a sequence with an unusually rugged
landscape, so we seek the value of the fluctuations $\d (b^2)^2$ in
the energetic variance per interaction $b^2$, as various sequences of
length $N$ are sampled. The conformationally averaged variance per
contact $b^2$ for a given sequence is given by $\sum_{i, j
= 1}^m \d \eps_{i j}^2 p_i p_j$ (see e.g. ref.~\cite{PandeVS97:bj}), where
$\d \eps_{i j}^2$ is the 
variance of an element of the pair interaction matrix between residue
types $i$ and $j$, e.g. the Miyazawa-Jernigan matrix~\cite{MiyazawaS96}. 

The average sequence to sequence fluctuation in $b^2$,
$\left< \d (b^2)^2 \right>$, is obtained by calculating the
fluctuations in the incorporation probabilities $p_i$, which decrease
with increasing chain length as $N^{-1}$. The result is 
\be
\left< \d (b^2)^2 \right> = \frac{4}{N} \sum_{i,j,k=1}^m \d \eij^2 \d
\eik^2 \, \pio \left( 1-\pio \right) p_j^o p_k^o \equiv \frac{\sb^2}{N}\: .
\label{varbsq}
\ee 
As chain length becomes large, $P(b^2)$ becomes an increasingly
sharply peaked function about the average $\left< b^2\right>$. We
apply the central limit theorem so that the probability $\PR$ a sequence has
ruggedness greater than say $\bg^2$ is just the integral of the tail
of a Gaussian, i.e. an error function. 
Typically the interaction matrix and pool
compositions are such that the thermal
energy of the heteropolymer is well above the ground state energy,
i.e. $\left<b^2\right>$ is sufficiently small that the temperature
$T>\Tg$ where $\Tg$ 
is the glass temperature in the random energy model (REM)~\cite{DerridaB81}.
Later we will be interested in special sequences that are
sufficiently rugged that without buffing they would be kinetically trapped
and thermodynamically stable in low
energy states at biological temperatures.  
Because of the $N$ dependence in equation~(\ref{varbsq}), along with
the fact that an error 
function has an exponential tail, such extra rugged sequences become
exponentially rare with increasing system size (see eq.~(\ref{eqbuff1})).

Given an ensemble of sequences with ruggedness $b^2$, we now seek the
fraction of these sequences with an energy gap larger 
than $\Del$. The relevant gap $\Del$ is between the energy of the
ground state conformation and 
the energy of the next lowest globally dissimilar structure. Hence the
distributions of state energies $P(E)$ can be well approximated by Gaussian
random variables of variance $N c b^2$, where $c$ is the number of
interactions per residue. If there are $\Om =
\e^{N s_o}$
total dissimilar conformations (all assumed compact with the same
$c$, here $s_o$ is the conformational entropy per residue), $\Om-1$ of them must be 
$\Del$ or higher in energy than a particular one. The particular
ground state can have any energy but is most likely to have energy
near the REM ground state for a collection of $\Om$ 
states~\cite{DerridaB81}.  Any 
of the $\Om$ states are candidates to be the ground state structure. 
Then the fraction of
sequences with gap larger than $\Del$ is given by the expression
\be
p(\Del |b) = \Om \int_{-\infty}^{\infty} d\!E  \, P(E) \, \left(
\int_{E+\Del}^{\infty} d\!E' \, P(E') \right)^{\Om-1} 
\approx \e^{-\sqrt{\frac{2 s_o}{c}} \, \frac{\Del}{b}}
\label{fractfunneled}
\ee
where the saddle point solution has been taken to obtain the last
expression, as worked out previously in ref.~\cite{SaliA94:jmb}. 
Note that the smoother the landscape (smaller $b$), the smaller the
fraction of sequences gapped to $\Del$. We assume here no selection in 
ruggedness for funneled sequences, i.e. funneled sequences have the typical
variance $\left< b^2 \right>$, but have an atypical density of
states. \\

{\bf Nucleation for trap escape.} $\;$ The alternative of buffed energy landscapes is not allowed in a
strictly mean-field picture of folding dynamics, where the landscape
is determined by the global statistical properties of the
mean-field solution. However, for larger proteins, folding and trap
escape depend on the spatially local (and/or local in sequence)
properties of the energy landscape~\cite{Wolynes97cap}. For folding,
these capillarity models resemble those of an ordinary first order
phase transition. On the other hand,
in capillarity models of the glassy dynamics for trap escape based on
the theory of random first-order phase transitions, fluctuations in
the height of escape barriers scale in the same way as the typical
barrier height~\cite{Kirkpatrick89}. This motivates an estimate for 
the size of the ensemble of sequences with anomalously small escape
barriers.

For a given sequence, the density of states is parameterized by two
important energy scales, the variance $b^2$, and the energy gap
$\Del$.~\footnote{Trap escape affects the prefactor to folding on a
gapped landscape, or the approach to the ground state on a buffed
landscape, so the relevant free energy profile only involves the
variance $b^2$.}  There are other features of the energy landscape
however, not completely captured by the density of states, such as the
barrier height $\Fdag$ between dissimilar configurations for a given
sequence. In mean field theory the characteristic barrier height
$\Fbardag$ is largely determined from the density of states and the
Hamiltonian. However in capillarity theory, the fluctuations in
barrier heights are large, and $\Fdag$ can deviate considerably from
$\Fbardag$.

Given a 
trap of energy $E$, the typical profile for growth of a nucleus involved in
trap escape is given by formula~(\ref{fnucleus}) with $F(0)$ replaced
by the energy of the trap
\be
\Fbar(\Ne) = E - f \Ne + \s \Ne^z \: .
\label{nucescape}
\ee
Given a composition with $b$,
the  free energy $F(N)$ of the ensemble  of untrapped structures is given 
by the total free energy $F(b,T)$ at temperature $T$. The value of
$F(b,T)$ depends on a critical  parameter 
\be
\bg = T \sqrt{2 s_o/c}
\label{bg}
\ee
which is the value of $b$ where the system is sufficiently rugged to be
glassy at biological temperature $T$. In terms of the total energetic
variance $B^2
\equiv N c b^2$ and the total conformational entropy  $S_o \equiv N s_o$, 
when $b < \bg$ the free energy 
$F(b,T) = -T S_o - B^2/(2 T)$ and when $b > \bg$, 
$F(b,T)$ equals the ground state energy $- \sqrt{2 S_o B^2}$. 
Equations~(\ref{nucescape}) and the free energy $F(b,T)$  together
determine the bulk free energy gain $f$.

We approximate the surface tension as proportional to the
estimate for the mean-field escape barrier per residue calculated by
us earlier~\cite{WangPlot97} on the
correlated landscape.  That calculation shows the surface
tension is always intensive 
and vanishes above a critical energy $\Estar$ or below a critical
ruggedness $\bstar$. \\

{\bf Atypical Buffed sequences.} $\;$ Eq.~(\ref{nucescape}) is
understood to be the typical 
profile for trap escape for a given overall composition. However 
fluctuations from this mean profile for various sequences will occur.
We expect that these fluctuations can be relatively large, since
reconfigurational barriers scale as $N^{1/2}$, but the distribution of a
random process
after $N$ events of residues joining the growing phase has a width
$\sim N^{1/2}$. 
Just as we found rare sequences that were anomalously rugged or
funneled, we are interested now in finding rare sequences that are
anomalously buffed, with low kinetic barriers.

Since the variance in interaction energies is $b^2$, when the amount
of new phase increases in a nucleation 
process by  $\d \Ne =1$ the change in free energy $\delta F$  for a given
sequence is chosen from a Gaussian distribution with mean $\d
\overline{F}(\Ne)$ and variance $c b^2$.
A particular nucleation process corresponds to a sequence of
increments $\{ \frac{\D F}{\D \Ne}(\Ne=1),\frac{\D F}{\D \Ne}(2),
\ldots ,\frac{\D F}{\D \Ne}(N)\}$,  
so that the probability $P\left[F(\Ne)\right]$ of a particular free
energy profile $F(\Ne)$ is given by a path integral over $\Ne$.
The probability of a free energy profile is analogous to the
probability amplitude of finding a particular path for a quantum
particle.  To 
find the probability $\phi(\Fdag)$ that a nucleation path has an
escape barrier from a given trap lower than some value $\Fdag$, and
that does not allow intermediate traps deeper than $\Fdag$ below
$F(b,T)$, we introduce an infinite square-well potential
$V\left(F(\Ne)|\Fdag\right)$ to constrain the
profile. $V\left(F(\Ne)|\Fdag\right)$ is zero as long as $F(b,T) -
\Fdag < F(\Ne) < E + \Fdag$, otherwise it is set to $\infty$. It
serves as an absorbing boundary for any paths wandering higher than
$\Fdag$ above $E$ or lower than $\Fdag$ below $F(b,T)$.
The probability we seek can then be expressed using a Green's function:
\be
G(F(b,T),N|E,0) = 
\int {\cal D}F(\Ne) \;\:
\mbox{e}^{-\int_0^N d\Ne \left[\frac{1}{2 c b^2} \left(\frac{\D F}{\D
\Ne}- \frac{\D \Fbar(\Ne)}{\D \Ne} \right)^2 -
V\left(F(\Ne)|\Fdag\right)\right]}\;
\delta\left(F(0)=E\right) \delta\left(F(N) = F(b,t)\right)
\label{Gprop}
\ee
where the delta functions ensure that the paths must start and end at the
trapped and untrapped free energies respectively.

To calculate the appropriate fraction of nucleation
paths $\phi(\Fdag)$ that have 
escape barrier from a particular trap  no larger than 
$\Fdag_{\mbox{\tiny{MAX}}} = E+ \Fdag - (F(b,T)-\Fdag)$, 
we must normalize by the free propagator
$G_{\mbox{\tiny{F}}}(F(b,T),N|E,0)$ in the absence 
of absorbing walls, i.e. with $V$ set to zero:
\be
\phi(\Fdag) = G(F(b,T),N
|E,0)/G_{\mbox{\tiny{F}}}(F(b,T),N|E,0)  \: .
\label{f1}
\ee

Evaluating these Green's functions is made easier by using the quantum
mechanical analogy.  We recognize the term $\D\Fbar/\D\Ne$ in
eq.~(\ref{Gprop}) as a time-dependent gauge transformation in one
dimension, so we see that the propagator to free energy $F$ after
$\Ne$ steps from the path integral problem satisfies an
imaginary-'time' Schrodinger equation
\be
\frac{\partial G}{\partial \Ne}(F,\Ne) - \frac{c b^2}{2}\frac{\partial^2
G}{\partial F^2} (F,\Ne) + \dot{\Fbar}(\Ne)\frac{\partial G}{\partial
F}(F,\Ne) + V(F(\Ne)|\Fdag) \;
G(F,\Ne) = \delta(F - E)\delta(\Ne)
\label{seq3}
\ee
where $\dot{\Fbar} \equiv \D \Fbar/\D \Ne = -f + z \s
\Ne^{-\left(1-z\right)}$. The situation is shown in Figure~\ref{fig1}.
Here $-i \Ne$ plays the role of time, $F$ the role of position,
$\hbar=1$ and mass $m=1/cb^2$.

The free propagator 
$G_{\mbox{\tiny{F}}}(F(b,T),N|E,0)$ is straightforward:
\be
G_{\mbox{\tiny{F}}}(F(b,T),N|E,0) = (2\pi c b^2 N )^{-1/2}
\, \e^{- \left(F(b,T)-\Fbar(N)\right)^2/2 c b^2 N}
\label{solnfreeprop}
\ee
where $\Fbar(\Ne)$ is again the mean-field potential. 

We have been unable to obtain an exact, closed form solution for the
non-separable, ``time-dependent'' problem with boundaries. However we
give in the appendix several analytical solutions in some reasonable limiting
cases, and solve the problem numerically in general.

There are polynomially many nucleation sites for a contiguous
nucleus, which can be estimated geometrically. We equate this with the number of 
possible ways or routes $N_{\mbox{\tiny{R}}}$ to escape from
a trap. If any of these nucleation sites are buffed to $\Fdag$, the
trap is then assumed to be buffed to $\Fdag$.
We assume independent buffing probabilities for each of the distinct
routes. Then the probability that $n$ of the routes are buffed is simply a
binomial distribution for $n$ events, the probability of each 
of which is $\phi(\Fdag)$ given in
equation~(\ref{f1}). Then the probability that at least one route is buffed
to $\Fdag$, i.e. the probability that a given trap of energy $E$ is buffed, is
$p_{\mbox{\tiny{B}}}\left(\Fdag|E,b\right) = 1 - \left( 1 -
\phi(\Fdag)\right)^{N_{\mbox{\tiny{R}}}}$. \\

{\bf Density of states and density of traps.} $\;$ To determine if a
sequence, rather than just a specific trap, is 
buffed we need to know how many 
candidate traps have energy $E$. We estimate the number of
candidate traps using the mean field calculation for a
``typical'' sequence. Then the effects of fluctuations sequence to
sequence can be investigated.

A configurational state
is a local trap if all states kinetically connected to it by a single local
move are higher in energy.  
Since the number of states connected to a given state equals $N \nu$
when each residue can move to $\nu$ other states on average, there is
a polynomial fraction of traps on uncorrelated
landscapes~\cite{BryngelsonJD89}, $f_{\mbox{\tiny{T}}}
\approx 1/N\nu$. On a correlated landscape however, there is an
exponentially small fraction of traps. To see this, let
the conditional probability that a state has energy $E'$ given that
it shares a 
fraction of $q$ contacts with a state of energy $E$ be $P_q (E' | E
)$. Then the fraction of states at energy $E$ that are traps is given
by
\be f_{\mbox{\tiny{T}}} (E) = \left( \int_{E}^{\infty} dE' \, P_q (E' | E)
\right)^{N\nu}\: ,
\ee 
with $q$ suitably chosen as below.

For a Hamiltonian consisting of interacting sets of $p$ contacts, the
conditional probability $P_q (E' | E)$
obeys a Gaussian distribution centered on $q^p \, E$ with variance 
$2 B^2 ( 1 - q^{2 p} )$~\cite{PlotkinSS02:quartrev2}. 
Thus when $q\rightarrow 1$, $P_q \left(E'
|E\right) \rightarrow \delta\left(E'-E\right)$, and when
$q\rightarrow 0$, or $p\rightarrow \infty$, the states become
uncorrelated and $P_q\left(E' |E\right) \rightarrow
P(E')$.   When $p=1$, the landscape statistics of a two-body
Hamiltonian are recovered. The more ``many-body'' the Hamiltonian is, the
more de-correlated states of a given structural similarity are.

Since we are looking at states connected by single kinetic moves, we
take $E'=E +\dE$ and $q=1-\dq$, with $\dE \sim {\cal O}(N)$ but small
compared to $E$, and $\dq \sim {\cal O} (N^{-1})$, since we envision
local moves of an intensive number of residues. Then 
the distribution of $\dE$ becomes a Gaussian with mean $p E \,\dq$ and
variance $4 B^2 p \,\dq$,
and the fraction of traps becomes
$
f_{\mbox{\tiny{T}}} (E) = [ (1/2)\; \mbox{erfc} (
\sqrt{p \, \dq} \,E/2 B )]^{N\nu}
$
where $\mbox{erfc}(x)$ is the complementary error
function. Thus $f_{\mbox{\tiny{T}}}$ is of the form of a fraction raised
to a large power. It is not significant until the argument or erfc (which is
intensive) is fairly large and negative. Then the error function may
be asymptotically expanded around one to yield
\be
f_{\mbox{\tiny{T}}} (E) \simeq \e^{-c(E)\, N}
\label{eqfracexp}
\ee
where 
$c(E) = \nu (\pi \sqrt{2} \, p\, \dq
E^2/B^2)^{-1/2} \exp( - p\, \dq E^2/2\sqrt{2} B^2)
$
is a function of an intensive argument. Retracing the steps in the
above argument 
for an uncorrelated landscape (or letting $p\rightarrow \infty$)
results in nearly every state 
constituting a trap, consistent with previous results.

The total number of traps $\Omt (E, b)$ at energy $E$ is then the
total number of 
states $\Omega$ times the fraction $P(E)$ of those states at energy
$E$, times the fraction $f_{\mbox{\tiny{T}}} (E)$ of those states that
are traps. A plot of the log number of states and log number of traps
is shown in figure~\ref{fignumtraps}.

The probability that all the traps at energy $E$ will be buffed to $\Fdag$ is 
\be
\PB \left(\Fdag , E, b\right) = 
p_{\mbox{\tiny{B}}}(\Fdag|E,b)^{\Omt (E,b) } \: ,
\label{eqPBE}
\ee
assuming independent buffing probabilities for each trap. This
assumption is
likely to underestimate $\PB$. On the other hand we have assumed
independent probabilities for each of the $N_{\mbox{\tiny{R}}}$ routes
out of a given trap to be buffed. This overestimates the value of
$p_{\mbox{\tiny{B}}}$, bringing it closer to unity. 

As energy increases above that of the ground state, the quantity $\PB$
rapidly approaches a value of unity. It is hardest to buff out the
lowest energies, even considering that there are more traps at higher
energies that must be buffed out. The fraction of sequences with
available traps buffed at {\it all}  energies that may be thermally
occupied is
\be
\PB \left(\Fdag, b\right) = \prod_{\Eg}^{0} \PB \left(\Fdag , E,
b\right) = \exp \left[ \int_{\Eg}^{0} \! dE \: \Omt (E,b) \, \ln
\left(p_{\mbox{\tiny{B}}}(\Fdag|E,b) \right) \right] \: .
\label{PBall} 
\ee

{\bf Sequences with buffed landscapes.} $\;$ Let the system have
non-extensive gap $\Del$, and be rugged enough so 
that the ground state is thermodynamically stable at the temperature
of the earth, $b > \bg$ in~(\ref{bg}) (Fig.~\ref{buffschem} shows a
schematic of buffing between low energy states).
Note that all the thermodynamic properties of buffed landscapes are
still the same as for random sequences. For example,
configurational states still have variances in energies, so there is
little or no thermal entropy left in the system when
$b>\bg$. It is the kinetic properties of the system which are
different.
Because the ground state is competing with polynomially other low
energy states, its Boltzmann weight is given by
$ \wn \approx (1+ c' N^{{}^\gamma} \: \e^{-\Del/T})^{-1}$
where $c'$ is a constant of ${\cal O}(1)$, and $\gamma$ is an exponent
less than one. So even for a non-extensive gap $\Del \sim {\cal
O}(N^{1/2})$, the system will still have large Boltzmann weight in the
ground state.

How rare are these buffed sequences, as system size grows larger?
Using eq.~(\ref{f1}) and the results from the Appendix, we find an
asymptotic expression for 
the fraction of surviving paths. To leading order 
\be 
\phi \sim
N^{1/2} \; \e^{-\frac{\pi^2 c b^2}{8 F^{\neq \; 2}} N - \; \frac{\s^2
(1-z)^2}{2 c b^2} N^{2 z -1} } \: .
\label{eq:phi}
\ee
The prefactor arises from the free propagator. In the exponent, the
first term is a  
diffusion term that corresponds physically  to the fact that buffed
sequences are rare because of the likelihood of the specific free
energy profile to fluctuate outside of $\pm \Fdag$ after $N$ steps.
The second term in the exponent comes from the 
forcing term in the diffusion equation and embodies
the fact
that buffed sequences are rare because typical sequences follow the
average free energy profile. However, perhaps unexpectedly, the
forcing term is subdominant, 
scaling more weakly with $N$ (recall $z$ is between $1/2$ and $2/3$
since the nucleus is roughened).
Since $\phi$ is exponentially small, the probability
a trap is buffed over {\em any} route (in the expression for
$p_{\mbox{\tiny{B}}}\left(\Fdag|E,b\right)$ above)
has only polynomial corrections: $p_{\mbox{\tiny{B}}} \simeq N \phi$. 

Because it is hardest to buff out the lowest energies, a sequence can
be said to be buffed when a band of energies starting at the ground
state and scaling polynomially with
system size have all traps buffed to $\Fdag$.
Buffing out polynomially many competing ground states gives
from~(\ref{PBall}) 
$\PB \approx p_{\mbox{\tiny{B}}} \e^{a N^{\a}}$ with $\a < 1$.
This again modifies the 
scaling by only by subdominant corrections. 
From eq.~(\ref{eq:phi}), more rugged sequences are harder to buff. The
fraction of buffed sequences stable in their ground state at
biological temperatures is thus dominated by those sequences having
$b^2 =\bg^2$ in eq.~(\ref{bg}). The fraction of sequences $\PR$ having
this ruggedness is given through eq.~(\ref{varbsq}). The total
fraction of buffed sequences $\Pbuff$ is then given by $\PR(\bg) \PB(\Fdag,
\bg)$ or 
\be
\Pbuff \sim \e^{- \frac{( 2 s_o T^2 - \langle
b^2\rangle )^2}{2 \sb^2} N - \frac{\pi^2 s_o}{4 F^{\neq \; 2}} N}
\label{eqbuff1}
\ee
For numerical values of the pair interaction energies such as those in the
Miyazawa-Jernigan matrix, and naturally occurring amino acid
abundances, sequence-to-sequence fluctuations of the barrier (second term  
in the exponent in eq.~(\ref{eqbuff1})) dominate the possibility of
finding a buffed foldable sequence, with the probability of finding a
sufficiently rugged composition playing a secondary role. 

From eq.s~(\ref{fractfunneled}) and~(\ref{eqbuff1}), we see that the
relative rareness of buffed vs. funneled sequences becomes a
quantitative issue. We address this in figure~\ref{pbufffun}, where
the fraction of sequences buffed to $\Fdag = 4 \kB T$ is plotted as a
function of chain length $N$, as well as the fraction of sequences
sufficiently gapped so that the forward folding barrier is $4 \kB
T$. The folding barrier is determined from a capillarity model,
however the gap must still scale extensively with chain length to
ensure a constant ratio of $\Tf/\Tg$. There is a crossover in the figure
beyond which it becomes easier to funnel. However for short sequences,
as well as for longer sequences with degrees of freedom removed (due
e.g. to native secondary or tertiary structure formation), buffing
becomes more likely.

{\bf Discussion.} $\;$ With increasing chain length
it becomes exponentially 
rare for sequences to be
significantly buffed. Likewise it is exponentially difficult to be
funneled, so the question is which exponential wins. Our estimates
suggest that funneling overall is the dominant mechanism at least for
larger proteins 
which fold on biological time scales. 
Still the buffing mechanism
raises some interesting practical questions. 

Is it possible to design and synthesize a ``buffed foldable protein''?
Despite their rareness, buffed proteins probably exist and would be
interesting objects, since they would possess a polynomial number of
accessible states that could act as memories.  Unfortunately, unlike
funneled landscapes, the definition of buffed landscapes does not
immediately lead to a design algorithm. We note that only recently has
the funneling recipe been explicitly used in achieving foldability
{\it de novo} in the laboratory~\cite{JinW03}. Still, combinatorial strategies are
possible. 
Although natural protein landscapes are most likely funneled, the buffing
mechanism may play a partial role in folding kinetics. Once a
significant part of the protein is at least locally fixed by the
funneling mechanism, the entropy is reduced to a level where buffing
can occur. This can act to remove specific intermediates or high
energy transition states on the landscape. 

Other mechanisms in the spirit of buffing may exist which can
facilitate folding. One can envision ``anti-buffed'' or ``gated''
landscapes which have regions of phase space closed off by high energy
barriers, so that the system folds faster by exploring a smaller
region of phase space, and avoiding regions which may contain deep
traps. The fraction of sequences having gated landscapes may be
worked out in a similar fashion as the fraction of buffed landscapes,
and is a topic for future work.  Such a corralling mechanism on a
funneled landscape is one way of describing the gatekeeper residues
discovered by Otzen and Oliveberg in
S6~\cite{OtzenD99:pnas}. Gating is similar in spirit to
Levinthal's original resolution to the kinetic paradoxes in protein
folding~\cite{LevinthalC69}, as well as issues which arise when
considering kinetic partitioning mechanisms~\cite{ThirumalaiD95}. 
Gating may also play a role in avoiding misfolded protein structures
responsible for aggregation-related diseases.

Funneling is a sufficient and likely but not necessary condition for
achieving foldability.  However the above folding mechanisms are not
mutually exclusive- funneled proteins may also exhibit buffing. The
extent to which buffing occurs in real proteins may be tested for
example by
looking at the mutational sensitivity of the weights of non-native
transient folding intermediates. Two candidates for these studies are
the $\alpha$-helical early folding intermediate in
$\beta$-lactalbumin~\cite{KuwataK01}, and the major kinetic traps in
$\alpha$-lactalbumin, whose populations may be measured by disulfide
scrambling~\cite{ChangJY02}.

The ways evolution has assisted
the folding and function of proteins are manifold- we have described
an alternative to funneling here. It will be interesting to see
what other organizing principles of energy landscapes are important in
describing structural and functional biology. \\

\noindent{{\bf Acknowledgments:} S.S.P. acknowledges funding from NSERC
and the Canada Research Chairs (CRC) program. P.G.W. acknowledges
funding from NIH grant R01-GM44557. \\

{\bf Appendix: Solutions for Green's function for buffing.} $\;$ In
this section we find approximate solutions to eq.~(\ref{seq3}), in
order to 
derive analytical expressions for $\phi(\Fdag)$ in eq.~(\ref{f1}). 
We are interested here in finding the fraction of sequences whose
kinetic barriers are below $\Fdag$, where $\Fdag \sim {\cal O}(\kboltz
T)$. The wave packet solution to~(\ref{seq3}) spreads out diffusively
to width $\Fdag$ in a number of steps $\Ne^{\ast} \sim
F^{\neq\: 2}/b^2$. But since both $\Fdag$ and $b$ are $\sim {\cal O}(\kboltz
T)$, $\Ne^{\ast} \sim {\cal O}(1)$. Hence we expect that by $N$
steps the wave function will have long since relaxed to the ground state. 
This also means that the 'time'-rate of change of the drift term in
eq.~(\ref{seq3}) is slow compared to the rate at which the wave packet
relaxes. The adiabatic parameter is given here by $|\left< 2 |\D H/\D
\Ne | 1 \right> /(E_2 - E_1 )^2 |$, where the numerator is the matrix
element of $\dot{\Fbar}(\Ne) \: (\partial G/\partial F)$ between the ground
and first excited eigenfunctions of the unperturbed Hamiltonian, and
the denominator is the difference in the first excited and ground
state eigenvalues. A straightforward calculation gives the parameter
as $L^3 \s /\pi^4 b^4 \Ne^{2-z}$ (where
$L \equiv \Fdag_{\mbox{\tiny{MAX}}}$ is given above eq.~(\ref{f1}) and is of
order a few $\kB T$). This
parameter is much smaller than unity over the whole range of $\Ne$
for all energies contributing to barriers to be buffed. Hence we
conclude that here the adiabatic approximation is a good one.
Then the time-dependent prefactor $\Fdotbar (\Ne)$ in~(\ref{seq3}) may
then 
be treated as a constant parameter, and the solution then given in
terms of that parameter. 

The adiabatic
approximation is a special case of a more general transformation. We
may eliminate the drift term 
in~(\ref{seq3}) by letting
$G(F,\Ne) = K(F,\Ne) \; \exp (g(F,\Ne))$, with
$g(F,\Ne) = \Fdotbar (\Ne) F/(c b^2) - (1/2 c b^2)
\int_{0}^{\Ne} \! d\Ne' \;\: \Fdotbar (\Ne')^2$.
This yields a new differential equation for $K(F,\Ne)$ with a
time and position-dependent sink term:
\be
\partial_{\Ne} K (F,\Ne) - (c
b^2/2) \partial_{\mbox{\tiny{FF}}} K (F,\Ne)
+ \left( (\ddot{\Fbar} (\Ne)/c b^2) F + V(F(\Ne)|\Fdag) \right)
K (F,\Ne) 
= \delta(\Ne) \delta(F - E) \lim_{\Ne \rightarrow 0}
\e^{-\frac{\Fdotbar(\Ne) \, E}{2 c b^2}} \: .
\label{seq4}
\ee

Since in the adiabatic approximation, the coefficient of the drift
term could be treated as a constant, 
the first term in
the parentheses in~(\ref{seq4}) can be neglected, since it is the
time-derivative of the drift coefficient. Then for $\Ne >0$,
$K(F,\Ne)$ satisfies a simple diffusion equation 
with absorbing walls, so that $K(F(b,T)-\Fdag, \Ne) = K(E+\Fdag,\Ne) =
0$, and the initial condition is $K(F,\Ne) = \d (F-E)$.  Thus the expansion
in eigenfunctions may readily be written down.
As we mentioned above, ground state dominance is an excellent
approximation for $\Ne >1$, so the solution is then given by the ground
state term in the eigenfunction expansion:
\be
G(F,\Ne) \cong (2/L) \: \e^{\frac{\Fdotbar(\Ne)}{c b^2} (F-E)}
\sin (\pi (E +\d F^{\neq} )/L) \sin ( \pi (F + \d F^{\neq})/L) \: \e^{-\left[ \frac{
\pi^2 c b^2}{2 L^2}  
+ \frac{\Fdotbar(\Ne)^2}{2 c b^2} \right] \Ne} \: ,
\label{Ggnd}
\ee
where $\d F^{\neq}
\equiv \Fdag-F(b,T)$.  
The propagator to $(F(b,T),N)$ is then $G(F(b,T),N)$. The fraction of
surviving paths is then given by $\phi(\Fdag)$ in~(\ref{f1}) along
with eq.~(\ref{Ggnd}) and the free propagator
in~(\ref{solnfreeprop}).  

The $N$-scaling of the fraction of buffed sequences may be checked with
an exactly solvable model. For near ground states buffed to $\Fdag$,
the width $L= 2\Fdag$ does not scale with $N$, and so can be replaced
by a parabolic potential of a fixed, effective width. 
Then $V(F(\Ne)|\Fdag)$ in eq.~(\ref{seq4}) is replaced by $\frac{1}{2}
\w^2 F^2 / c b^2$, where the effective frequency $\w$ determines the
spring stiffness. 
Here the differential equation~(\ref{seq4}) is equivalent to a
Schrodinger equation with $\hbar = 1$, time $t$ replaced by $-i
\Ne$, position replaced by free energy $F$, and particle mass replaced
by $1/c b^2$. Shifting coordinates so the effective harmonic well is
centered at zero, the solution to eq.~(\ref{seq4}) may be obtained
exactly by integrating the Lagrangian over all
paths~\cite{FeynmanRP65}. The result is: 
\be
K_{\mbox{\tiny{HO}}} \left( F, \Ne | F_o , 0\right) = \left(
\w^{-1} \: 2 \pi c b^2 \sinh (\w \Ne)\right)^{-1/2}
\e^{S_{\mbox{\tiny{CL}}} (F,\Ne|F_o,0)}
\ee
where
in the quantum mechanical analogy $S_{\mbox{\tiny{CL}}}$ represents
the (extremal) action along the classical path. Calculating this for
large $N$ 
shows that, analogous to the ``forcing term'' we saw before, the
classical action $S_{\mbox{\tiny{CL}}} ( F, N | F_o , 0)$ does not dominate
the scaling behavior. Instead the prefactor, measuring deviations from
the path of least action, plays the most important role,
scaling as $\sim \e^{-\w N /2}$ for large $N$. Following the same arguments as
before, this is essentially the scaling law for the fraction of buffed
sequences. Comparing the scaling law here with that in
eq.~(\ref{eqbuff1}), we see that the effective frequency $\w$ is such
that the unforced propagators in the hard-wall and parabolic-wall
cases have the same survival probability after $N$ steps. \\

{\bf  Figure Captions} \\

{\small FIG.~\ref{fig1} $\;\;$ Nucleation free energy profiles in trap
escape, as a function of the number of escaped residues $\Ne$. Shown
are the mean free energy profile (dashed line), and a typical profile
for sequences constrained to have barriers smaller than $E+\Fdag -
(F(b,T)-\Fdag)$. Also shown schematically is the propagator $G(F,\Ne)$
for various values of $\Ne$. The system rapidly relaxes to ground
state wavefunction after ${\cal O} (1)$ steps (see text).\\
}

{\small FIG.~\ref{fignumtraps} $\;\;$ Long dashed line: The log number
of states vs. energy. 
Solid line:
Approximate log number of traps from equation~(\ref{eqfracexp}).
Short dashed
line: Log number of traps for the error function expression of
$f_{\mbox{\tiny{T}}}$ in the text above eq.~(\ref{eqfracexp}).
The approximation is accurate for energies near 
the ground state, but falls off too rapidly for energies near
zero. There are still traps at $E=0$ because states may have positive
energies as well. Estimates are used here for the number of states per
residue ($\Omega = \e^{N \times 2}$), system size ($N=100$), and
number of connected states ($\nu=1$).  Pair interactions are taken
($p=1$), energies are in units of $\sqrt{N} B$, and $\dq = 1/N$.\\
}

{\small FIG.~\ref{fignumtraps} $\;\;$ Schematic of the energy
landscape for a sequence with buffed ground states, projected onto a
configurational coordinate. The barriers between the ground states are
reduced to $\Fdag$, but the overall variance of energy between states
is not reduced. The lowest kinetic barriers between the states
determines the escape rate. Along the  coordinate(s) where the landscape
is buffed, the kinetic barriers between states are reduced.
\\
}

{\small FIG.~\ref{pbufffun} $\;\;$ The fraction of foldable sequences,
on a log scale, as a function of chain length $N$. Both funneling and
buffing mechanisms are shown here. Dashed line: Fraction of funneled
sequences with forward folding barrier $\Fdag=4\kB T$, and
$\Tf/\Tg=1.6$. Solid line: Fraction of sequences buffed to $\Fdag=4\kB
T$. The adiabatic approximation for the Green's function in eq.~(\ref{seq3})
is used (see appendix). The crossover from buffing to funneling
suggests the possibility of 
a compound mechanism for generating a functional protein sequence. The
funneling mechanism removes most of the entropy while guiding the
protein to a smaller ensemble of similar structures. Then for this
reduced collection of states, dynamics between individual traps is
most likely mediated by buffing. This dynamics may be related to
functionally important motions in proteins~\cite{FrauenfelderH91}.
For funneled sequences we assume no ruggedness selection
occurs. We scale the MJ interaction parameters so that in our units
$\Tf/\Tg \approx 1.6$, a common value taken from the
literature~\cite{OnuchicJN95:pnas}. Buffed sequences must be selected
to be sufficiently rugged to be stable at biological temperatures, and
must have low kinetic barriers to be accessible on biological time
scales.  (INSET) Comparison of the fraction of buffed sequences from
the adiabatic approximation and from full numerical solution to
eq.~(\ref{seq3}). 
Several values of ruggedness are plotted. Circle: b=1.255, Triangle: b=1.4, 
Square: b=2.0. 
\\
}

\begin{figure}[htb]
\hspace{.7cm}
\psfig{file=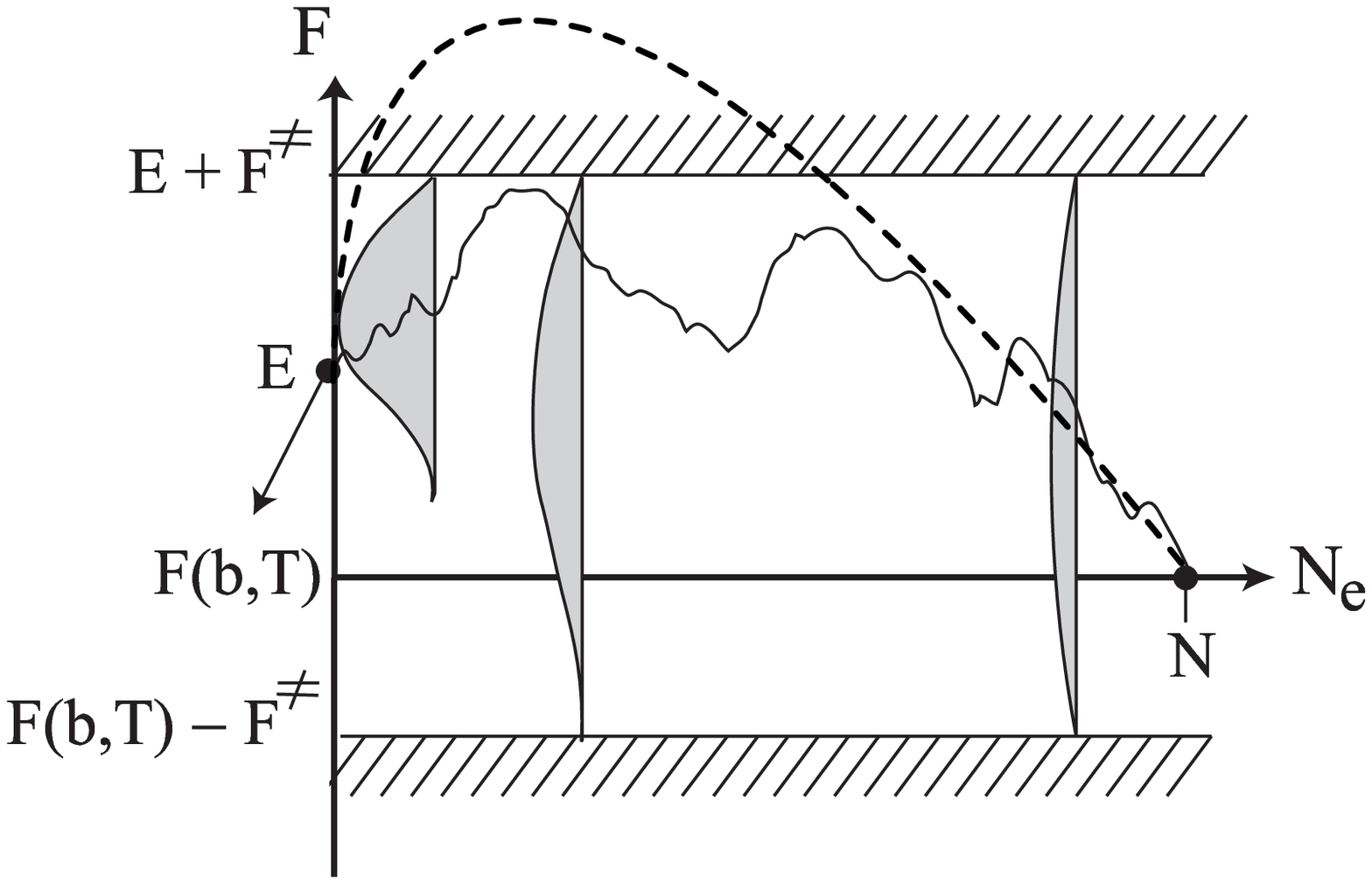,height=4cm,width=6cm,angle=0}
\caption{}
\label{fig1}
\end{figure}

\begin{figure}[htb]
\hspace{.7cm}
\psfig{file=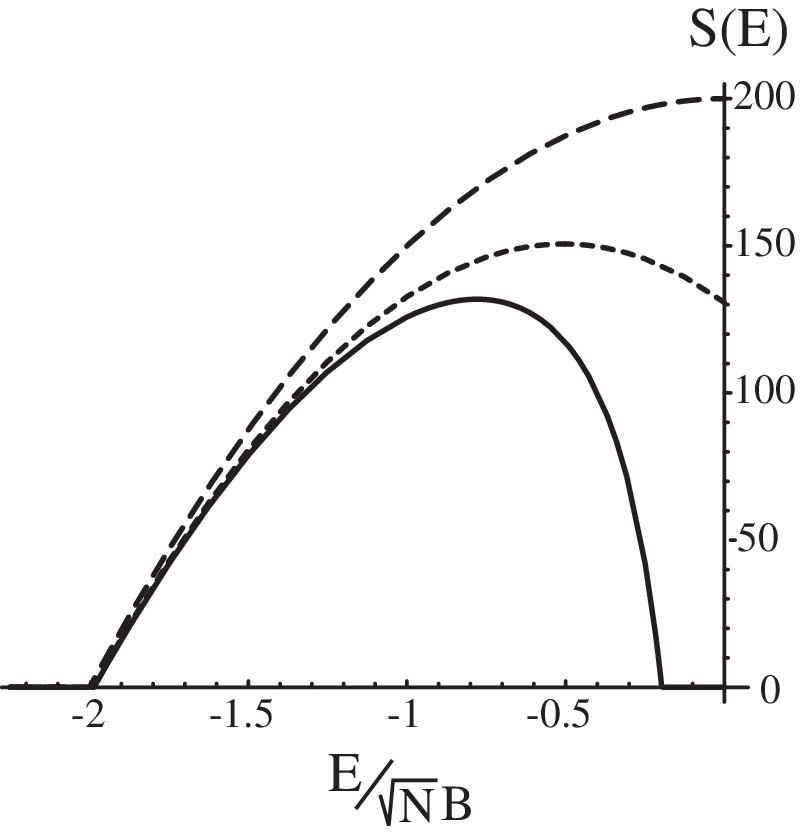,height=4cm,width=4.5cm,angle=0}
\caption{}
\label{fignumtraps}
\end{figure}

\begin{figure}[htb]
\hspace{.7cm}
\psfig{file=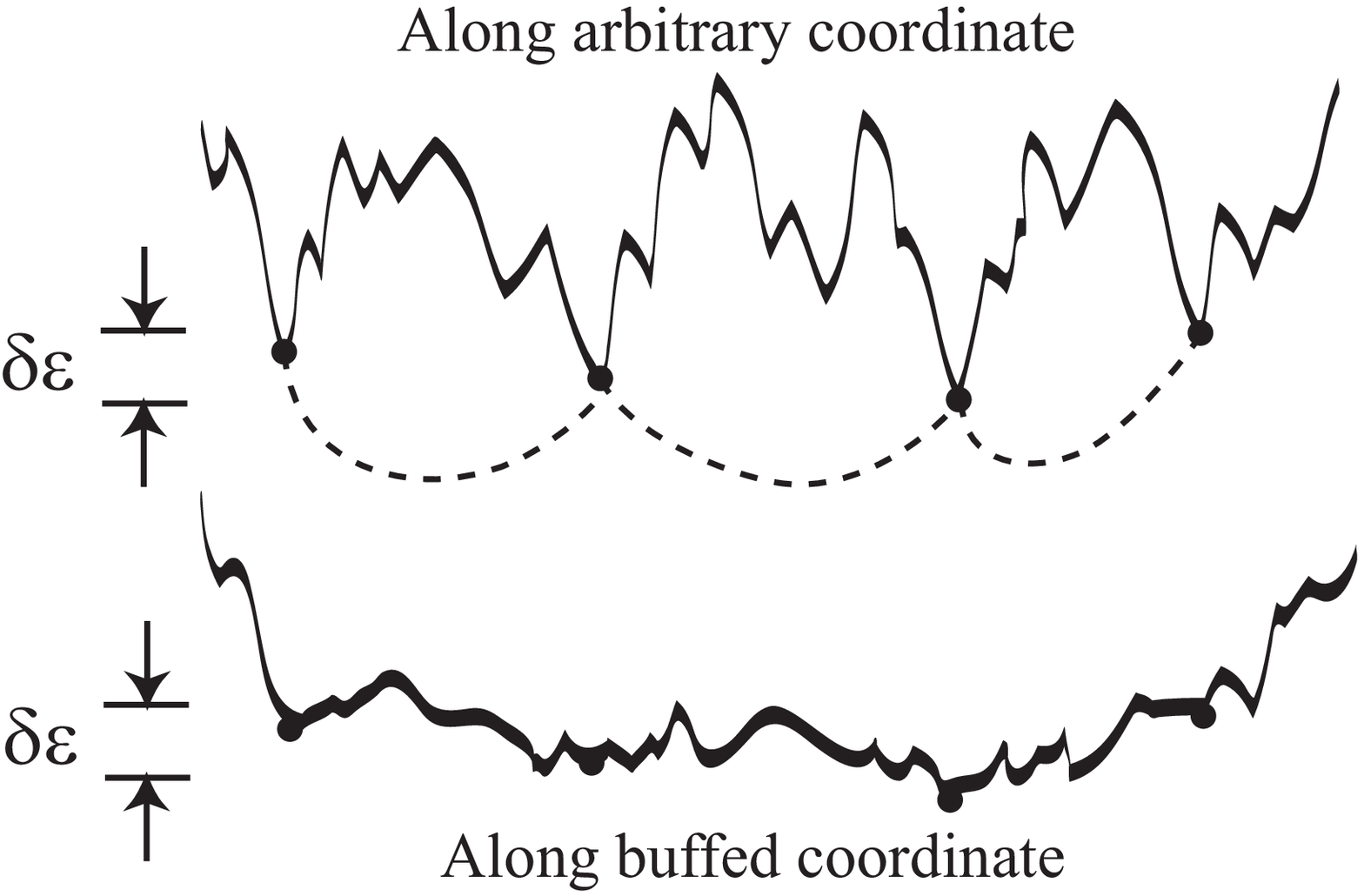,height=4cm,width=6cm,angle=0}
\caption{}
\label{buffschem}
\end{figure}

\begin{figure}[htb]
\hspace{.7cm}
\psfig{file=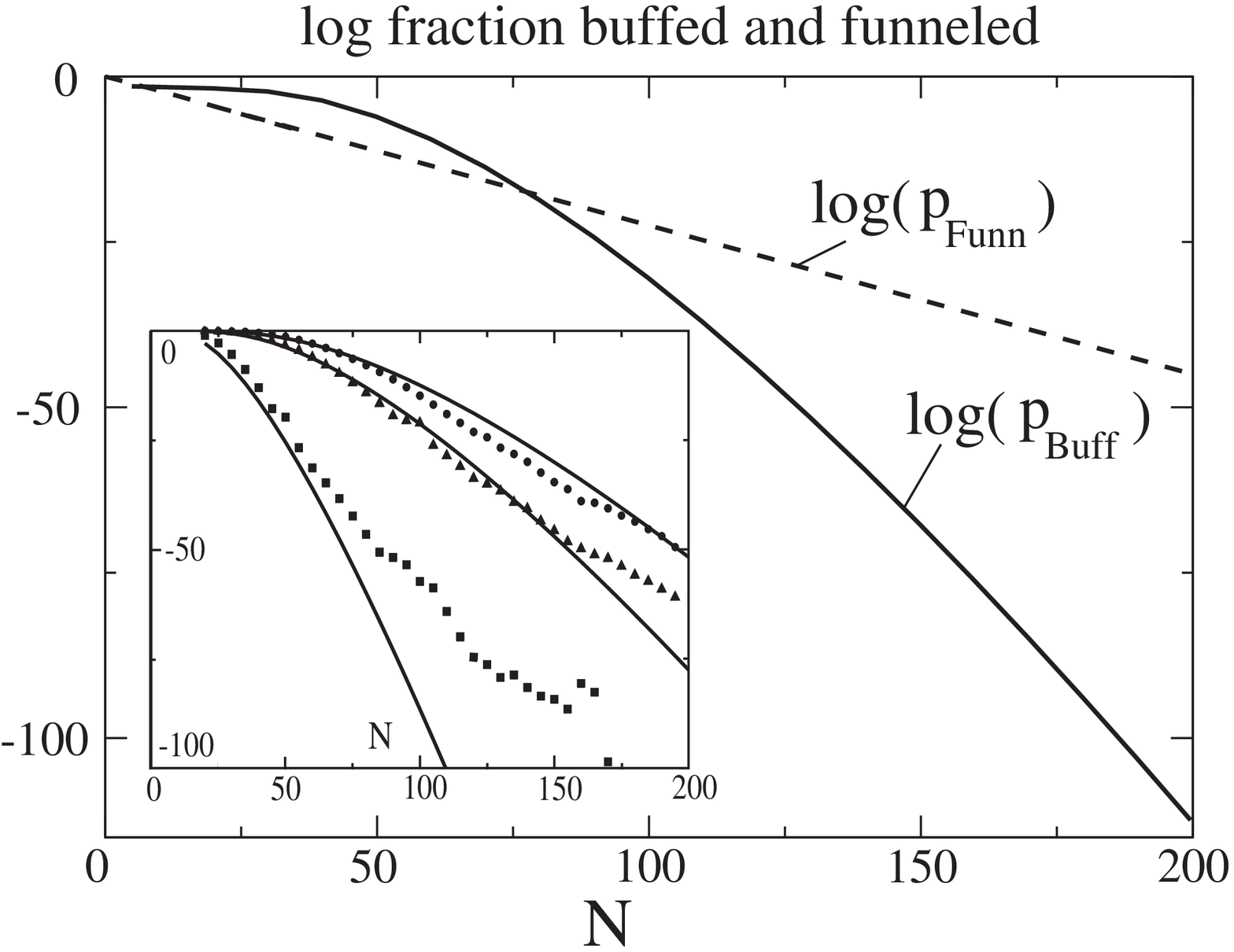,height=6cm,width=7.5cm,angle=0}
\caption{}
\label{pbufffun}
\end{figure}

\pagestyle{myheadings}
\addcontentsline{toc}{section}{References}

\end{document}